\documentclass{PoS}
\def\be{\begin{equation}}
\def\ee{\end{equation}}
\usepackage{graphicx}

\title{A survey of large $N$ continuum phase transitions}

\ShortTitle{A survey of large $N$ continuum phase transitions}

\author{\speaker{Rajamani Narayanan}\\
        Department of Physics, Florida International
University, Miami, FL 33199\\
        E-mail: \email{rajamani.narayanan@fiu.edu}}

\author{Herbert Neuberger\\
        Department of Physics and Astronomy, Rutgers University,
Piscataway, NJ 08855\\
        E-mail: \email{neuberg@physics.rutgers.edu}}

\abstract{
The main focus of this talk is the
physics of large $N$ QCD on a continuum torus. 
A cascade of phase transitions associated with the breaking
of $U(1)$ symmetries will be discussed. 
The continuum Wilson loop as a function of
its area will be discussed along with its universality properties
and the associated double scaling limit. Some recent progress
in twisted Eguchi-Kawai is presented. Gauge field topology
and $\theta$ vacuua are also discussed in the context of
large $N$ gauge theories.
Phase transitions in 2D large $N$ principal chiral models 
are compared with similar transitions in large $N$ gauge theories.
Finally, connections to some topics in string theory and
gravity are briefly described.}

\FullConference{The XXV International Symposium on Lattice Field Theory\\
		 July 30-4 August 2007\\
		 Regensburg, Germany}

\begin{document}

\section{Large $N$ QCD in the 't Hooft limit}

Large $N$ gauge theories~\cite{'t Hooft:1973jz}
are qualitatively similar to
QCD with three colors and it is long held hope 
to solve it analytically for $N=\infty$.
Only planar diagrams contribute in this limit and
fermions in the fundamental representation are
naturally quenched as long as the number of flavors is
finite. Researchers in string theory and gravity also address the
problem of large $N$ gauge theories but they are a long way
from solving it analytically~\cite{Aharony:1999ti}.

We will discuss various physical properties of large $N$
QCD in the 't Hooft limit. We will start with well known
results in $d=2$ and proceed to a discussion of relatively
new results in $d=3$ and $d=4$. The theory will be regulated
using the lattice formalism and
the lattice bare coupling,
$b=\frac{1}{g^2N}$, will be held fixed as
$g\rightarrow 0$ and $N\rightarrow\infty$.
We will assume that we only have a finite number
of fermion flavors and therefore fermions will be
naturally quenched as long as we are in the confined
phase and there is no chemical potential.
The continuum limit corresponds to $b\rightarrow\infty$.
All our discussion will be on
a periodic lattice at a  finite physical volume. Specifically,
\begin{itemize}
\item{$d=2$: $l_{x,y}=\frac{L_{x,y}}{\sqrt{b}}$.
The physical sizes $l_x \le l_y$ are kept fixed as 
$L_{x,y}$ and $b$ are taken to $\infty$.}
\item{$d=3$: $l_{x,y,z}=\frac{L_{x,y,z}}{b_I}$;\newline
We will use the tadpole improved coupling~\cite{Lepage:1996jw}, 
$b_I=b e(b)$
where $e(b)$ is the average value of the plaquette.
The physical sizes
$l_x \le l_y \le l_z$ are kept fixed as $L_{x,y,z}$ and $b$ are 
taken to $\infty$.}
\item{$d=4$: $l_{x,y,z,t}=L_{x,y,z,t}
\left[\frac{48\pi^2b_I}{11}\right]^{\frac{51}{121}}
e^{-\frac{24\pi^2b_I}{11}}$.
The physical sizes
$l_x \le l_y \le l_z \le l_t$ are kept fixed as $L_{x,y,z,t}$ 
and $b$ are 
taken to $\infty$.}
\end{itemize}

\section{$U(1)$ symmetry and continuum reduction}

The lattice gauge action for $SU(N)$ gauge theory
on a $L_1\times\cdots L_d$ periodic lattice 
is given by
\begin{eqnarray}
S=\frac{bN}{2}\sum_{n,\mu\ne\nu} Tr[ U_{\mu,\nu}(n)
+U_{\mu,\nu}^\dagger (n) ] \\
U_{\mu,\nu}(n)=U_\mu (n) U_\nu (n+\mu) U_\mu^\dagger (n+\nu) 
U_\nu^\dagger (n).
\end{eqnarray}
In addition to the local gauge symmetry, the above action
has a $Z^d_N$ global symmetry under which
the Polyakov loop in the $d$ directions get rotated
by a $Z_N$ phase factor:
\be
U_\mu(n) \to e^{i2\pi k_\mu/N} U_\mu(n) \ \ 
{\rm for}\ \  n_\mu=L_\mu\ \ {\rm and}\ \ 
 0\le n_\nu < L_\nu
\ \ {\rm for}\ \  \mu\ne\nu
\ \ \ 0\le k_\mu < N
\label{znt}
\ee
Each $Z_N$ becomes a $U(1)$ in the $N\rightarrow\infty$ limit.

If the $U(1)$ symmetry is not broken in a given direction
on a $L_1\times\cdots L_d$ lattice at a fixed coupling $b$,
then no physical quantity depends on the size of that direction.
The proof of the above statement is a simple extension of
the original Eguchi and Kawai~\cite{Eguchi:1982nm}
argument to a $L_1\times\cdots L_d$
lattice. The continuum limit of the above statement 
(namely, $L_\mu\to\infty$, $b\to\infty$, such that
the physical size $l_\mu$ is kept fixed) 
is referred to
as continuum reduction~\cite{Narayanan:2003fc}.

If continuum reduction holds in a certain direction,
the parallel transporter in that direction can be folded
using periodic boundary conditions to construct a transporter
of arbitrary length. This enables one
to consider Wilson loops of arbitrary size on a finite box.

Continuum reduction has interesting consequences for fermions.
First one notes that the gauge
transformations can be extended from $SU(N)$
to U(N) and still get the same result
for fermionic gauge invariant quantities. One can then convert
(\ref{znt}) to
\be
U_\mu(n) \to e^{i\frac{2\pi k_\mu}{NL_\mu}} U_\mu(n);\ \ \ 
0\le k_\mu < N
\ee
using a U(N) gauge transformation that obeys periodic
boundary conditions. The gauge field action is invariant
under the above transformation and therefore an observable
made out of a single fermion (like the quark condensate)
cannot depend upon $k$. But a fermionic observable that
is made out of one quark and a different anti-quark 
(like a $\pi^+$ meson)
will depend upon $(k-q)$ if one quark sees a gauge field
with $k_\mu$
and the other quark sees a gauge field with $q_\mu$. 
Therefore,
one can have continuous momenta in this direction
where the discete momentum interval 
on the lattice
are filled by $\frac{2\pi(k-q)_\mu}{NL_\mu}$. This is called the quenched
momentum prescription~\cite{Gross:1982at} for mesons.

\section{Large $N$ QCD in two dimensions}
 The two $U(1)$ symmetries remain unbroken for all values of
$b$ and $L_{x,y}$~\cite{Bhanot:1982sh}.
Therefore
the problem 
can be reduced to a single site on the
lattice with $U_1$ and $U_2$
being the two $SU(N)$ degrees of freedom. 
There is no dependence on the box size $l_x$ or $l_y$
for any $0\le l_x \le l_y \le \infty$. 
Large $N$ $QCD$ in $d=2$ is always in the confined phase and there
is no dependence on the temperature.

\subsection{Gross-Witten transition}
The plaquette operator is $P=U_1 U_2 U^\dagger_1 U^\dagger_2$
and its eigenvalues $e^{i\theta_p^j}$, $j=1,\cdots,N$ are gauge
invariant. Consider
the eigenvalue distribution, $\rho(\theta_p;b)$, for
$-\pi < \theta_p \le \pi$ obtained upon averaging over
$U_{1,2}$ using the Wilson action. 
Gross and Witten~\cite{Gross:1980he}
showed that this observable
 exhibits non-analytic
behavior as a function of $b$.
\begin{eqnarray}
\rho(\theta_p;b) =& 
{\frac{2b}{\pi}}\cos{\frac{\theta_p}{2}}
\sqrt{\frac{1}{2b} - \sin^2\frac{\theta_p}{2}},&
b \ge\frac{1}{2};\ \ |\theta_p| < 2 \sin^{-1}\sqrt{\frac{1}{2b}}\cr
\rho(\theta_p;b) =& \frac{1}{2\pi}\left(1+2b\cos\theta_p\right), &
b\le \frac{1}{2};
\ \ |\theta_p|\le\pi
\end{eqnarray}
The eigenvalue distribution has a gap for $b>\frac{1}{2}$
and it does not have a gap for $b<\frac{1}{2}$.
The lattice theory has a 
third order phase transition at $b=\frac{1}{2}$
and this transition is referred to as the Gross-Witten transition.
This transition is a lattice phenomenon since the location
of the transition does not scale with the lattice size and
the continuum
theory is only in the phase where the 
eigenvalue distribution of the plaquette operator has a gap.

\subsection{Wilson loops in large $N$ $2d$ QCD\label{wil2d}}

The plaquette operator is a Wilson loop whose area goes
to zero as one goes to the continuum limit. For
a physical loop, consider
a rectangular Wilson loop of size $n\times m$.
The Wilson loop operator can be obtained by folding on a single
site lattice and is
given by $W(n,m) = U_x^nU_y^m \left(U_y^mU_x^n\right)^\dagger$.
Let $t=\frac{n m}{2b}$ be the parameter that
characterizes the dimensional area. Consider a continuum
Wilson loop of a fixed area by taking $b\rightarrow\infty$,
$nm\rightarrow\infty$ while keeping $t$ fixed. 
Since, ${\rm Tr} W(n,m) 
= \left[{\rm Tr}W(1,1)\right]^{nm}$~\cite{Gross:1980he},
it is easy to show that
${\rm Tr} W(t) = e^{-\frac{t}{2}}$.

One can proceed further and get analytical expressions
for ${\rm Tr} W^n(t)$. In this context, it is useful to
consider
the generating function 
$F(z,t)=\frac{1}{2}+\sum_{n=1}^\infty \frac{\left<{\rm Tr} W^n(t)
\right>}{z^n}$
where $z$ 
is a complex variable.
Then,
$F(z,t)$ satisfies~\cite{Durhuus:1980nb}
\be
z=\frac{2F(z,t)+1}{2F(z,t)-1}e^{-tF(z,t)}
\label{doeqn}
\ee
and the expectation value of powers of Wilson loops 
are~\cite{Bassetto:1999dg,Kazakov:1980zi,Rossi:1980ms,Gross:1994mr}
\be
\left<{\rm Tr} W^n(t)\right> = \frac{1}{n} L^{(1)}_{n-1}(nt)
e^{-\frac{nt}{2}} \label{wil2de}
\ee
The expectation value of the
distribution of the eigenvalues, $e^{i\theta}$, of $W$
is given by 
\be
\rho(\theta,t)=-\frac{1}{\pi}{\rm Re}
F(e^{i\theta},t)
\label{dorho}
\ee

\subsubsection{Critical behavior of Wilson loops}
The above results imply a 
critical behavior of Wilson loops as a function of area.
The expectation value of
arbitrary powers of Wilson loops,
$\left< {\rm Tr} W^n(t)\right>$, as given
by (\ref{wil2de}) are analytic functions of $t$.
Yet, the eigenvalue distribution, $\rho(\theta,t)$,
exhibits a non-analytic behavior as a function 
of $t$~\cite{Durhuus:1980nb}.
One way to see this is to ask what $\rho(\pi,t)$ is
as a function of $t$. Setting $z=-1$ in (\ref{doeqn})
results in
\be
2F(-1,t)=\tanh\frac{tF(-1,t)}{2}
\ee
$F(-1,t)=0$ is always a solution but the non-zero
solution for
$F(-1,t)$ when
$t>4$ is favored.
Therefore, the critical point is $t=4$ and
the distribution has a gap for $t<4$ and
does not have a gap for $t>4$.
It is also clear from the above equation that
$F(-1,t) \propto \sqrt{t-4}$ as $t\to 4$~\cite{Janik}.

A non-trivial critical behavior is observed if one
stays at the critical point, $t=4$, and asks for
the behavior of $\rho(\theta,4)$ close to $\theta=\pi$.
If we set $z=-e^{iy}$ and
$t=4$ in (\ref{doeqn}), one finds that
$\rho(y,4)\propto y^\frac{1}{3}$ and therefore the
number of eigenvalues in an arc of length $dy$, near $y=0$, is
proportional to $Ny^\frac{4}{3}$. Therefore, the level
spacing is proportional to $N^{-\frac{3}{4}}$~\cite{Janik}.

This physical transition in Wilson loops as a function of area
from weak coupling  ($t<4$) to strong coupling ($t>4$)
is called the Durhuus-Olesen transition.

\subsubsection{Double scaling limit of the Durhuus-Olesen 
transition\label{conj}}

The critical behavior of the Wilson loops as a function of area
results in a universal function in the double scaling limit
where one takes $t\to 4$ and $\theta\to\pi$.
The double scaling limit can be studied by considering
\be
O_N(y,b)=
\left(\frac{N}{12}\right)^{\frac{1}{4}} \sqrt{\frac{2\pi }{Nb}}
\frac{e^{\frac{N}{2b}}}{2^N} \left < \det \left( e^{\frac{y}{2}}
+e^{-\frac{y}{2}} W \right) \right>
\label{modeleqn}
\ee
$W=\prod_{j=1}^n U_j$ is a product of $n$ independently and
identically distributed $SU(N)$ matrices, $U_j=e^{i\epsilon H_j}$.
Each $H_j$ is traceless and its entries are independently
distributed Gaussian random variables. We let $\epsilon\to 0$
and $n\to \infty$ such that the area, 
$\frac{4}{b}=t=n\epsilon^2$,
is kept fixed.

One can
use an integral representation over
Grassmann fields and a perturbation expansion in $\epsilon$
to show that~\cite{nnprep}
\be
O_N(y,b)=\left(\frac{N}{12}\right)^{\frac{1}{4}}
\int d\rho e^{N\left[\ln\cosh\rho-\frac{b}{8}
(2\rho-y)^2\right]}.
\label{mock2d}
\ee
The double scaling limit amounts to taking $b\to 1$ and
$y\to 0$. Using the appropriate scaling exponents obtained
in the previous section,
we define scaled variables, 
$\xi$ and $\alpha$ by
\be
y=\left(\frac{4}{3N^3}\right)^{\frac{1}{4}}\xi;\ \ \ 
b=1+\frac{1}{\sqrt{3N}}\alpha.
\label{yarho}
\ee
An
expansion in $\frac{1}{\sqrt{N}}$,
results in the following Generalized Airy integral
as the universal scaling function:
\be
\lim_{N\rightarrow\infty} O_N(y,b)=
\zeta(\xi,\alpha)=\int du e^{-u^4 -\alpha u^2 + \xi u}
\label{airy}
\ee

The above equation describes the universal behavior in
the double scaling limit.
The conjecture of dimensional reduction is that
the universal function $\zeta(\xi,\alpha)$
defined in the double scaling limit
for
large $N$ QCD in $d=2$ is also 
obeyed by large $N$ QCD in $d=3$ and $d=4$.

\begin{figure}
\vspace{1.1cm}
\begin{center}
\centerline{\includegraphics[width=0.8\textwidth]{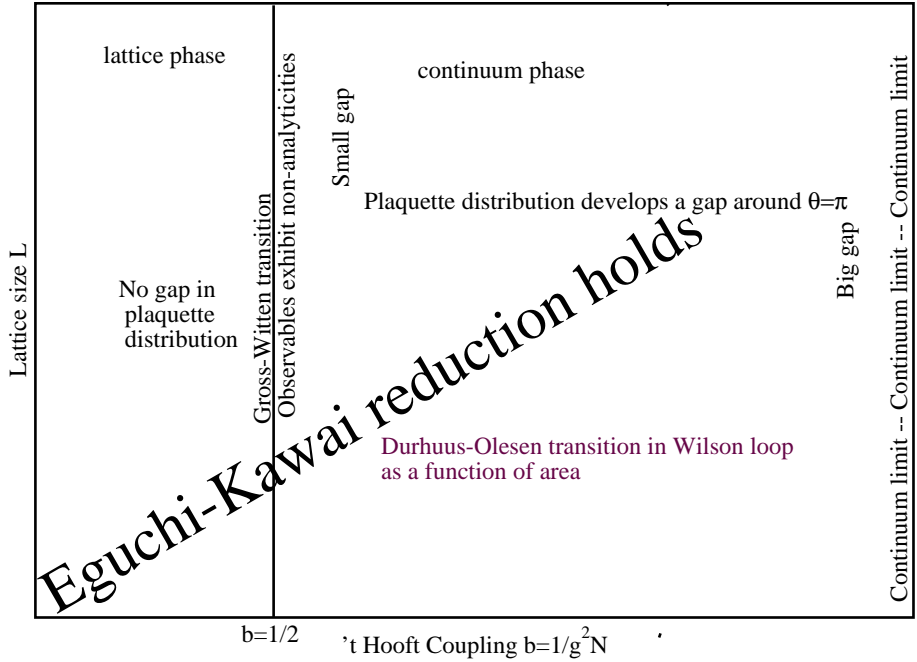}}
\caption{Summary of large $N$ QCD in $d=2$\label{twod}}
\end{center}
\end{figure}

We end this section with
a pictorial summary of large $N$ $2d$ QCD in Fig.~\ref{twod}

\section{Large $N$ QCD in three dimensions}

The $U^3(1)$ symmetries are spontaneously broken on finite lattices
and Eguchi-Kawai reduction does not hold. The continuum
theory can exist in several phases labeled as $0$c, $1$c,
$2$c and $3$c corresponding to the number of $U(1)$ symmetries
that are broken~\cite{Narayanan:2003fc,Bursa:2005tk,Narayanan:2007ug}. 
The theory is confined in the $0$c phase
and deconfined in the $1$c phase.

\subsection{Transition in the plaquette operator}

Like in $2d$,
the eigenvalue distribution of the plaquette operator undergoes
a transition from having no gap for small $b$ to having a gap
for large $b$~\cite{Bursa:2005tk}.
Numerical work indicates that this transition is
either second or third order and occurs at $b\approx 0.43$ 
for the Wilson gauge action.
Despite some similarities with the Gross-Witten transition,
there is no evidence that the transition
in $d=3$ is in the same universality class as Gross-Witten.
Like in $2d$, the location of the
transition does not scale with the lattice size
and therefore it is a lattice transition. 
Like in $2d$, the continuum
theory is always in the phase where the 
eigenvalue distribution of the plaquette operator has a gap.

\subsection{Setting the scale in $3d$ large $N$ QCD}
Although there is some progress in obtaining analytical
results in $3d$ QCD, much of the results are obtained
by numerical means. One example where numerical results
confront analytical results is the case of string tension.
Using the Hamiltonian formalism and a parameterization
of the spatial gauge potential using a complex
$SL(N,C)$ matrix, one can 
analytically obtain a good approximation to the vacuum wave-function and
thereby obtain a result for the
string tension~\cite{Karabali:1998yq}. The result for
the string tension is
\be
\sigma = \frac{1-\frac{1}{N^2}}{8\pi b^2}
\ee
for all $N$.

Lattice computations with
$N=2,3,4,5,6,8$ have been 
performed\cite{Bringoltz:2006gp,Bringoltz:2006zg}
 by computing the
correlation functions of Polyakov loops at zero spatial
momentum.
The Polyakov loops themselves are constructed
using smeared gauge fields. The numerical results
have very small statistical errors (typically less
than 0.4\%) and they 
deviate from the
above analytical result. 
The numerical result for the string tension is consistently 
smaller than the analytical result for all $N$ and
the differences are large for small $N$.
The extrapolation of the lattice results to $N\rightarrow\infty$
gives
\be
\sqrt{\sigma} b = 0.1975 \pm 0.0002 - 0.0005;
\ee
and this has to be compared with the analytical result, 
$\frac{1}{\sqrt{8\pi}}=0.19947114\cdots$
The first error in the numerical estimate is statistical
and the second error which is always negative comes from
performing two different fits (either a single cosh
or a double cosh, which takes into account the presence
of an excited state in the correlation function).
The difference between analytical and lattice results
although statistically significant is still quite small.

\subsection{Deconfinement transition in $3d$ large $N$ QCD}

Consider large $N$ QCD on a $L^3$ torus at a fixed lattice
coupling $b$. 
An order parameter suitable for studying the phase transitions
we are interested in is~\cite{Bhanot:1982sh}
\begin{eqnarray}
\bar P_{x,y,z} &=& \left < P_{x,y,z} \right > \cr
P_{x,y,z} &=& \frac{1}{2 L_x L_y L_z} \sum_n  1 - \left | \frac{1}{N} 
Tr {\cal P}_{x,y,z}(n)
\right |^2 \cr
{\cal P}_{x,y,z}(n) &=& \prod_{m=1}^{L_{x,y,z}} U_i(n+m\hat i).
\end{eqnarray}
The quantity $P_{x,y,z}$ takes values in the range $[0,0.5]$ on
any gauge field background and one chooses the 
$x$, $y$ and $z$ 
directions for each configuration
such that $P_x < P_y < P_z$.

Fixing $L$ and $N$, 
one finds the $b_1(L)$ such that one of
the three $U(1)$ symmetries is broken for $b > b_1(L)$:
$\bar P_{x,y,z}=\frac{1}{2}$ for $b < b_1(L)$ and
$\bar P_x < \frac{1}{2}$, $P_{y,z}=\frac{1}{2}$ for
$b > b_1(L)$. One finds that $b_1(L)$ is independent of $N$
for large enough $N$ ($N=47$ is usually sufficient).
Then one finds that the tadpole improved critical coupling,
${b_1}_I(L)$, scales with $L$ and $l_1 = L/{b_1}_I(L) = 5.90(47)$.
This shows that there are two phases in the continuum limit
and
$l_1$ defines a physical size such that 
\begin{itemize}
\item All three $U(1)$ symmetries are unbroken for $l > l_1$
and there is no dependence on $l$ in this phase ($0$c).
\item One of the three $U(1)$ symmetries is broken for $l < l_1$
and the theory depends on the size of the broken direction in
this phase ($1$c).
\end{itemize}
The critical size $l_1$ does not depend on $l_{y,z}$
since the $U(1)$ symmetries are not broken in the
$l_y$ or $l_z$ directions and continuum reduction
holds in those directions. Therefore, the system is in
$l_x\times \infty^2$ if $l_x < l_1$. It is natural to
identify the finite direction with that of temperature
in the deconfined phase.
Therefore, $0$c to $1$c is the deconfinement transition and
the deconfinement temperature is 
\be
\frac{t_c}{\sqrt{\sigma}}=\frac{1}{l_1\sqrt{\sigma}}=0.86(7)
\ee
Since continuum reduction holds in the $0$c phase in all
three directions,
there is no temperature dependence in physical quantities
in the $0$c phase.
A latent heat measurement is needed to directly establish
the order of the phase transition in the large $N$ limit.

Conventional numerical studies of $SU(N)$ in $d=3$ on
$L^2\times L_t$ with $L_t=3,4,5$ and $L$ as high as $48$
indicate the following.
Both SU(2) and SU(3) gauge theories exhibit a second order 
deconfinement transition~\cite{Engels:1996dz}.
The case of SU(4) is marginal and the transition is difficult to
establish~\cite{de Forcrand:2003wa}. 
The $Z_4$ spin model has continuously varying exponents.
Small $L_t$ indicate the transition is first order but larger
$L_t$ possibly indicate a second order transition.
The transition is clearly first order from 
SU(5) onwards~\cite{Holland:2005nd}.
The large $N$ limit obtained from extrapolating the
$N=4,5,6$ results~\cite{Liddle:2005qb} 
for the critical temperature is consistent
with the critical size for the $0$c to $1$c transition.

\subsection{Transition to large $N$ QCD in a small box}

Now consider large $N$ QCD in the $1$c phase
by picking a box of size
$L_x\times L_y\times L_z$
with $L_x \le L_y \le L_z$ and $b > b_1(L_x)$. 
The box size has been chosen such that
the $U(1)$ symmetry in the x-direction is broken.
As $b$ is increased,
the $U(1)$ in the $L_y$ direction will break
at some $b_2(L_y,L_x)$. For the special case
of $L_y=L_z$, one of the two $U(1)$ will break
and the broken direction will be called $L_y$.

The theory is in the $2$c phase for $b>b_2(L_y,L_x)$ and
the $2$c phase exists in the continuum theory since
${b_2}_I(L_y,L_x)$ scales with $L_y$.
There is a characteristic size associated with the
$1$c to $2$c transition, namely, $l_2(l_x)$,
obtained by taking the limit of $L_y/{b_2}_I(L_y,L_x)$
as $L_y$ goes to infinity while keeping $L_x/L_y$ fixed.
This critical
size does not depend on $l_z$ since $l_z\ge l_2(l_x)$
and the $U(1)$ symmetry in that direction is not broken.
Therefore, the system is in $l_x\times l_y\times \infty$
while in the $2$c phase.
It is natural to associate the two finite directions with
a small periodic box and the infinite direction as time.
One cannot address confinement in the $2$c phase since
only one direction has an infinite extent.

\begin{figure}
\vspace{1.1cm}
\begin{center}
\includegraphics[width=0.6\textwidth]{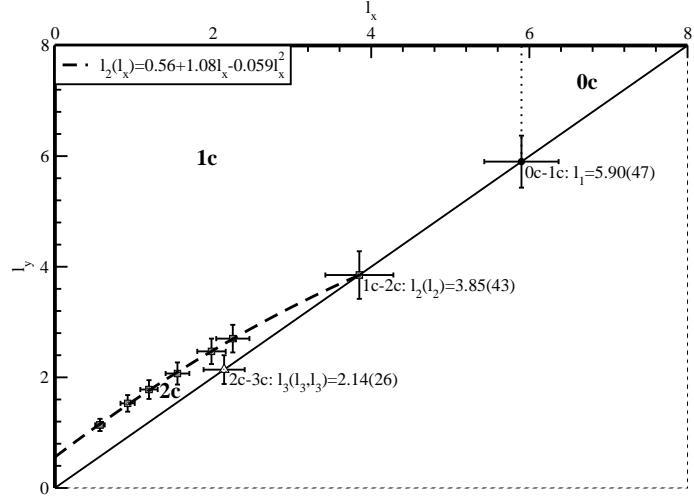}
\caption{Phase diagram in the $(l_x,l_y)$ plane for 
$l_x\le l_y \le l_z$\label{xvsy}}
\end{center}
\end{figure}

The system goes into the $2$c phase only if 
$l_x < l_y \le l_2=0.65(9)l_1$.
For
$0\le l_x \le l_2$,
the system goes into the $2$c phase
at some $l_2(l_x)$ 
with $l_2(l_2)=l_2$.
Numerical extrapolation of
$l_2(l_x)$ indicates that $l_2(0)>0$.
The full picture is shown in Fig.~\ref{xvsy}.
Note that the critical line connecting the $1$c and
$2$c phase is such that one can start in the
$2$c phase with $l_x=l_y < l_2$ and go into the $1$c phase
by keeping $l_y$ fixed and reducing $l_x$.
Finally,
rotational symmetry is present in the two broken directions
if $l_x=l_y < l_2$.

\subsection{Large $N$ QCD in a small box at high temperature}
Large $N$ QCD on a $L_x\times L_y\times L_z$ box 
with $b>b_2(L_x,L_y)$ undergoes
a phase transition at $b_3(L_x,L_y,L_z)$ beyond
which all three $U(1)$ symmetries are broken.
The system is in the $3$c phase for $b > b_3(L_x,L_y,L_z)$
and corresponds to large $N$ QCD in a small box at high
temperature.

It should be possible to do perturbation theory deep in
the $3$c phase but one has to account for the zero modes of
the gauge fields on the torus.
There are no zero modes to deal with
if one considers the theory on $S^2\times S^1$.
For a small
radius of $S^2$, one can show
using perturbation theory that the $U(1)$ symmetry associated
with $S^1$ is broken when the radius of $S^1$ gets smaller
than a certain size~\cite{Papadodimas:2006jd}.

Numerical computations show that ${b_3}_I(L,L,L)$ scales
properly with $L$ and therefore the continuum theory can also exist
in the $3$c phase.
$l_3(l_x,l_y)$ is the characteristic size associated
with the $2$c to $3$c phase transition and 
$l_3(l_3,l_3)=l_3=0.36(5)l_1$.

\subsection{Wilson loop operator in $d=3$ large $N$ QCD}

In order to test the proposed conjecture in section~\ref{conj}
we need
a definition of the Wilson loop operator that does
not suffer from perimeter divergence.
We know that
numerical computation of the string tension 
are performed by using correlators of smeared Polyakov loops.
Therefore,
we define smeared 
rectangular Wilson loops~\cite{Narayanan:2006rf} 
of size $n\times m$ by $W(n,m;f;k=\frac{(n+m)^2}{4};b)$.
The parameter $f$ is the APE smearing factor
and $k$ is the number of smearing iterations. $k$ should
be proportional to the square of the 
perimeter for dimensional reasons since
the smeared propagator is of the form~\cite{Bernard:1999kc}
\be
h_{\mu\nu}(q) = f(q) \left (\delta_{\mu\nu} - 
\frac{\tilde q_\mu\tilde q_\nu}{\tilde q^2}\right) + 
\frac{\tilde q_\mu\tilde q_\nu}{\tilde q^2}
\ee
\be
f(q) = e^{\frac{fk}{4}{\tilde q^2}}
\ee

\subsubsection{Test of the critical behavior of 
Wilson loops\label{wilcr}}

Wilson loops show critical behavior even 
without smearing~\cite{Bursa:2005tk}.
The distribution of the eigenvalues of the Wilson loop
operator show good agreement with the Durhuus-Olesen
distribution as given by (\ref{doeqn}) and (\ref{dorho}).

It is necessary to study the continuum limit of this
critical behavior and for this purpose 
one has to show that the eigenvalue distribution of
$W(n,m;f;k=\frac{(n+m)^2}{4};b;N)$ 
in the $0$c phase at a fixed $n,m$ and $f$
undergoes a transition
from having no gap at small $b$ to having a gap at large $b$
as $N\rightarrow\infty$.
Furthermore,
the critical $b_c(n,m;f;N)$ should scale properly 
as $N\rightarrow\infty$,
$nm\rightarrow\infty$ 
such that 
\be
\lim_{N\to\infty}\frac{b_c(n,m;f;N)}{\sqrt{nm}} = \frac{1}{l_w(f)}
\ee
has a finite limit.
$l_w(f)$ is the critical size of the Wilson loop and it
will depend upon $f$.

\begin{figure}
\vspace{1.1cm}
\begin{center}
\includegraphics[width=0.6\textwidth]{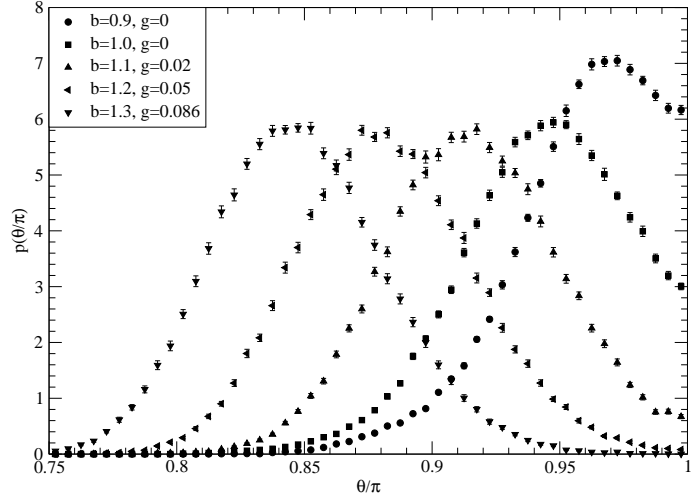}
\caption{Distribution of the eigenvalue closest to 
$\pi$\label{gapdist}}
\end{center}
\end{figure}

We fix the lattice size $L$ and size of the color group, $N$.
We then
pick one value of $f$ and pick a square Wilson loop,
$n=m$.
We study the eigenvalue distribution,
$e^{i\theta}$ of the Wilson loop
operator for a range of $b$ such that distribution goes
through the transition. This is illustrated in Fig.\ref{gapdist}
where the eigenvalue closest to $\pi$ is plotted as a function
of $b$ for a $6\times 6$ loop at $N=37$ on a $8^3$ lattice. 
The non-zero gap is estimated~\cite{Narayanan:2006rf} 
by matching the mean and
variance to the Tracy-Widom~\cite{tracy} distribution
for the universal distribution of the largest eigenvalue
in the Gaussian ensemble. The explicit equation for the
gap is
\be
g=1-\frac{1}{\pi}\left[
\left<\theta_N\right> + 1.96400484\sqrt{\left<\theta_N^2\right>
-\left<\theta_N\right>^2}\right]
\ee
where $\theta_N$ is the eigenvalue closest to $\pi$.

\subsubsection{Extracting the critical coupling and related
parameters}

We perform
a numerical calculation of the expectation value of
\be
O_N(r,b)=
\left< \det (e^{\frac{r}{2}}+e^{-\frac{r}{2}}W)\right>
\ee
where $W\in SU(N)$ is a $n\times n$ Wilson loop at a 
lattice coupling $b$. We expect this observable to
exhibit critical behavior at $b=b_c(n)$ and $r=0$
as $N\rightarrow\infty$.
If the double scaling limit is in the same universality
class as the one in $2d$
then we expect 
\be
\lim_{N\rightarrow\infty} {\cal N}(b,N)
O_N\left(r=
\left(\frac{4}{3N^3}\right)^{\frac{1}{4}}\frac{\xi}{a_1(n)},
b=b_c(n)+\frac{\alpha}{\sqrt{N}a_2(n)}\right) = \zeta(\xi,\alpha)
\label{match}
\ee
where ${\cal N}(b,N)$ is a normalization factor 
and $\zeta(\xi,\alpha)$ is the Generalized Airy integral
as given by (\ref{airy}).

A test of the above conjecture proceeds by first obtaining
an estimate for $b_c(n)$, $a_1(n)$ and
$a_2(n)$.
Since $\det W =1$, it follows that $O_N(r,b)$ is an even
function of $r$. It is also evident from (\ref{airy})
that $\zeta(\xi,\alpha)$ is an even function of $\xi$.
Let
\be
O_N(r,b)= C_0(b,N) + C_1(b,N) r^2 +  C_2(b,N) r^4 + \cdots
\label{taylor}
\ee
be the Talyor's series for $O_N(r,b)$.
Consider
\be
\Omega(b,N) = \frac{ C_0(b,N) C_2(b,N)}{C_1^2(b,N)}.
\ee
It is clear that this quantity will be the same
for $O_N(r,b)$ and ${\cal N}(b,N)O_N(r,b)$.
It is also clear that this quantity remains the same
if we replace $r$ by 
$\left(\frac{4}{3N^3}\right)^{\frac{1}{4}}\frac{\xi}{a_1(L_w)}$
and view the Taylor's series as an expansion in $\xi$.
Therefore, the value of this quantity is well defined
at the critical point, $b=b_c(n)$ and is given by
\be
\Omega(b_c(L_w,N),N) = 
\frac{\Gamma(\frac{5}{4}) \Gamma(\frac{1}{4})}{6 \Gamma^2(\frac{3}{4})} = \frac{\Gamma^4(\frac{1}{4})}{48\pi^2}= 0.364739936
\label{method1}
\ee
The first equality in (\ref{method1}) is obtained by
evaluating the same quantity starting from (\ref{airy}).
and we have used
\be
\int_{-\infty}^\infty du u^{2k} e^{-u^4} =\frac{1}{2}\Gamma
\left[\frac{2k+1}{4}\right]
\ee
Therefore, we obtain an estimate of $C_i(b,N)$; $i=0,1,2$,
using Montecarlo simulations and thereby obtain an
estimate of $\Omega(b,N)$. We then use (\ref{method1})
to obtain an estimate of $b_c(n)$ at a fixed $N$ and
extrapolate it to $N\to\infty$. 

The parameter $a_2(n)$ is defined via
\be
b=b_c(n) + \frac{\alpha}{a_2(n)\sqrt{N}}
\ee
Therefore, we use the following relation
\be
\frac{d\Omega(b,N)}{d\alpha}|_{\alpha=0}=
\frac{1}{a_2(n)\sqrt{N}} \frac{d\Omega}{db}|_{b=b_c(n)}
= \frac{\Gamma^2(\frac{1}{4})}{6\sqrt{2}\pi}
\left( \frac{\Gamma^4(\frac{1}{4})}{16\pi^2} -1\right)
=0.0464609668
\label{a2ex}
\ee
to obtain $a_2(n)$ at a fixed $N$. Since this is a
sub-leading quantity, errors are larger in this quantity
than in $b_c(n)$.

Upon substitution of
\be
r=
\left(\frac{4}{3N^3}\right)^{\frac{1}{4}}\frac{\xi}{a_1(n)}
\ee
in (\ref{taylor}), we conclude that
\be
\sqrt{\frac{4}{3N^3}} \frac{1}{a_1^2(n)}
\frac{C_1(b_c(n),N)}{C_0(b_c(n),N)}
= \frac{\pi}{\sqrt{2}\Gamma^2(\frac{1}{4})}=0.16899456
\label{a1ex}
\ee
and we use this relation to obtain an estimate
of $a_1(n)$. The results as a function of $N$ can be
extrapolated to get the value at $N=\infty$.

\begin{figure}
\vspace{1.1cm}
\begin{center}
\includegraphics[width=0.6\textwidth]{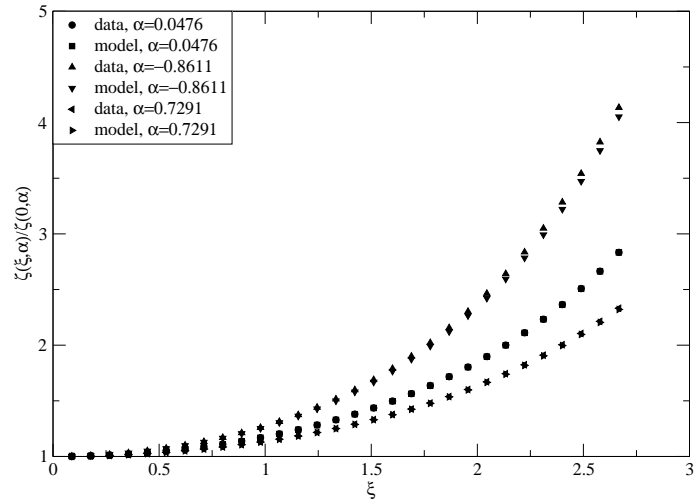}
\caption{The Generalized Airy integral
is compared to lattice data for a
$4\times 4$ Wilson loop at $N=47$ on $8^3$ lattice.
\label{scaling}}
\end{center}
\end{figure}

\begin{figure}
\begin{center}
\vspace{1.1cm}
\includegraphics[width=0.6\textwidth]{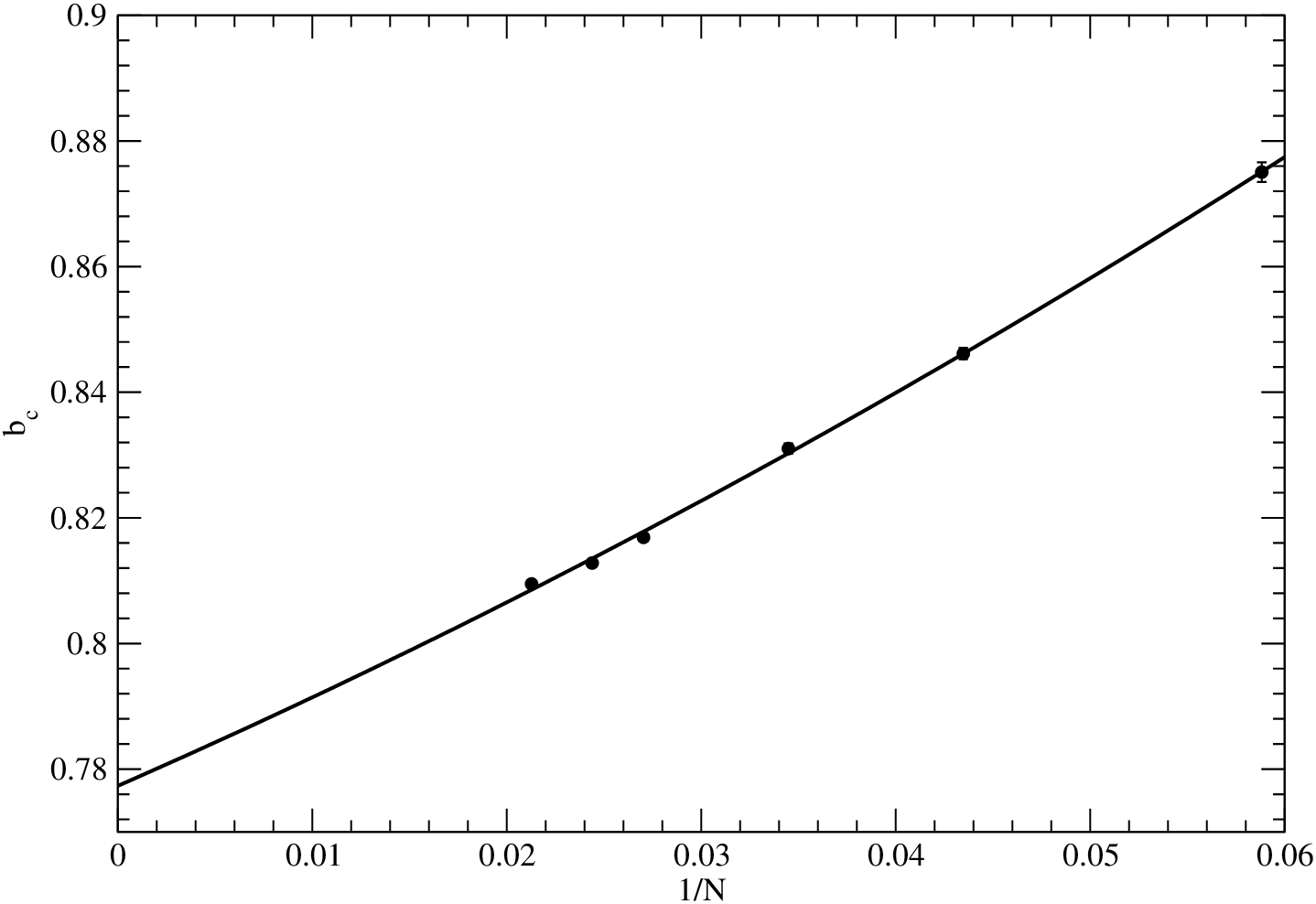}
\caption{Plot of $b_c(N)$ as a function of $\frac{1}{N}$
for $4X4$ loop on $8^3$ lattice.
\label{bc4}}
\end{center}
\end{figure}

\begin{figure}
\vspace{1.1cm}
\begin{center}
\includegraphics[width=0.6\textwidth]{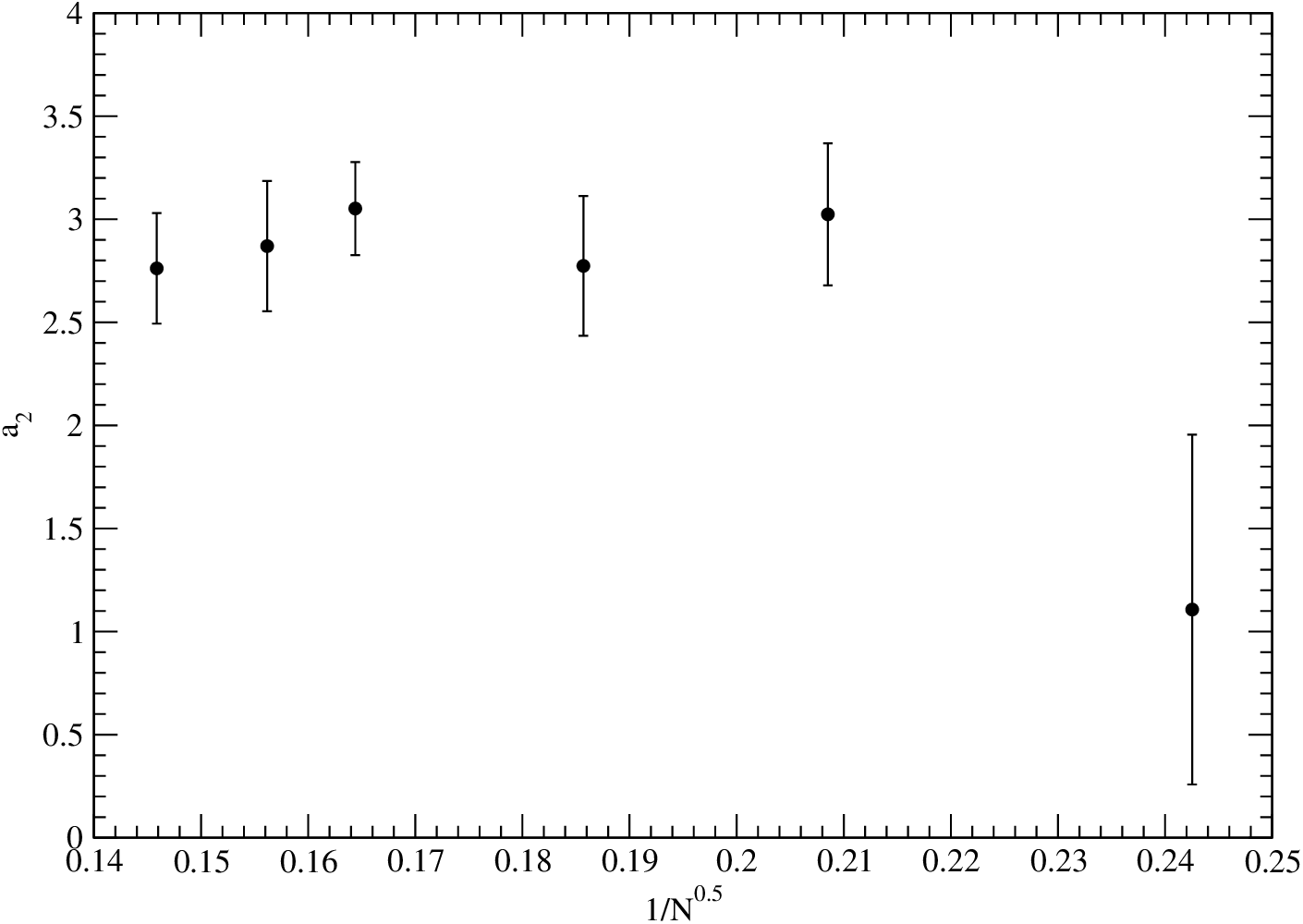}
\caption{Plot of $a_2(N)$ as a function of $\frac{1}{\sqrt{N}}$
for $4X4$ loop on $8^3$ lattice.
\label{a24}}
\end{center}
\end{figure}

\begin{figure}
\begin{center}
\vspace{1.1cm}
\includegraphics[width=0.6\textwidth]{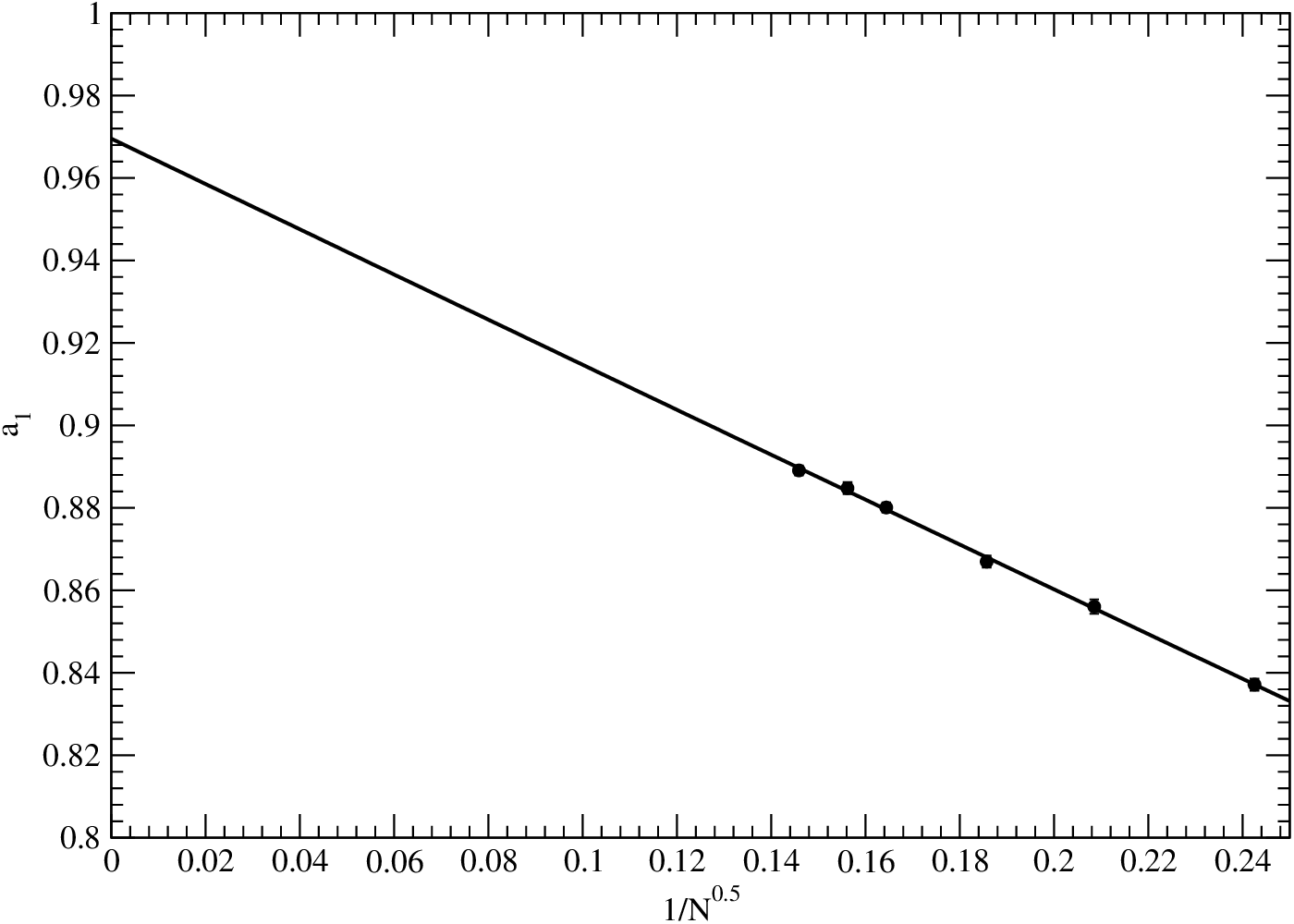}
\caption{Plot of $a_1(N)$ as a function of $\frac{1}{\sqrt{N}}$
for $4X4$ loop on $8^3$ lattice.
\label{a14}}
\end{center}
\end{figure}

As an example of the above procedure, one finds
$b_c=0.8095(4)$, $a_2=2.76(27)$ and $a_1=0.8891(12)$
for a $4\times 4$ Wilson loop at $N=47$ on a $8^3$
lattice with $f=0.03$. The resulting function on
the lattice as defined by (\ref{match}) matches
quite well with the Generalized Airy integral.
Due to the arbitrary normalization that is involved
in the matching, one possible way of checking the
agreement is to look at the ratio, 
$\frac{\zeta(\alpha,\xi)}{\zeta(\alpha,0)}$ for
several values of $\alpha$ as a function of $\xi$.
Fig.~\ref{scaling} shows such a comparison.
The approach to the large $N$ limit of $b_c$,
$a_2$ and $a_1$ are shown in Figs.~\ref{bc4}-\ref{a14}.
The agreement with the Generalized Airy integral gets
better as one gets closer to the large $N$ limit.

\begin{figure}
\vspace{1.1cm}
\begin{center}
\includegraphics[width=0.6\textwidth]{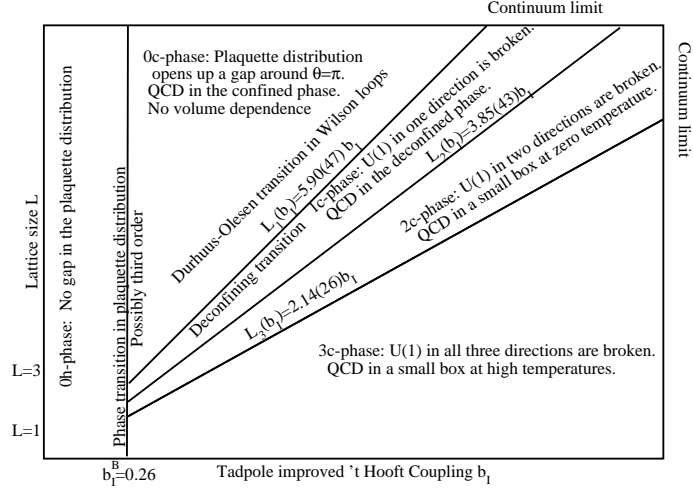}
\caption{Summary of large $N$ QCD in $d=3$ on $L^3$ 
lattice\label{threed}
}
\end{center}
\end{figure}

We end the discussion of large $N$ $3d$ QCD
with a pictorial summary shown in Fig.~\ref{threed}.

\section{Large $N$ $4d$ QCD}

Large $N$ $4d$ QCD was
reviewed in 
Lattice 2005~\cite{Narayanan:2005en}.
We begin by summarizing the current status and 
focus on topics that were not covered in the previous review.

There is a transition in the plaquette operator.
This occurs at $b=0.36$ for all $L^4\ge  4^4$ and the
transition is first order~\cite{Campostrini:1998zd,Kiskis:2002gr}.
Both $0$c phase and $1$c phase have a continuum
limit~\cite{Kiskis:2003rd}. 
The critical size scales according to
\be
L_c(b) = (0.250\pm0.025) \left ( \frac {11}{48\pi^2 b_I }\right )^
{\frac{51}{121}} e^{\frac{24\pi^2 b_I}{11}}
\ee
This transition is the deconfinement transition
studied on the lattice by taking the large $N$ limit
using $N=2,3,4,6,8$~\cite{Lucini:2005vg}.
The deconfinement transition is first order and
the latent heat has been measured through the jump
in the internal energy. The latent heat is
found to be $\delta \epsilon \approx
0.26 N^2 \epsilon_{\rm SB}$ where
$\epsilon_{\rm SB}$ is the blackbody energy density per
massless vector particle~\cite{Kiskis:2005hf}. 
Critical sizes associated with the
$1$c-$2$c, $2$c-$3$c and $3$c-$4$c transition have not
been determined yet.

\begin{figure}
\vspace{1.1cm}
\begin{center}
\includegraphics[width=0.6\textwidth]{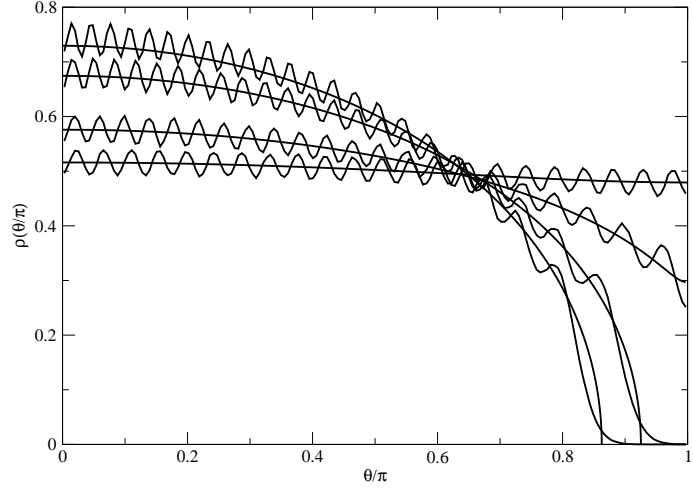}
\caption{Fit of the
lattice data to the Durhuus-Olesen distributions for four
different sizes of wilson loops, namely,
$l_wt_c=0.740,0.660,0.560,0.503$. The associated areas ($t$ in the
$2d$ notation)
that describe the continuous curves are
given by $t=8.06,4.60,2.82,2.30$ respectively.\label{fit4d}}
\end{center}
\end{figure}

Transitions in 
smeared Wilson loops were first studied in 
$4d$~\cite{Narayanan:2006rf} before
starting the careful investigation of the double
scaling limit in $3d$. The transition 
fits the Durhuus-Olesen behavior as shown in Fig.~\ref{fit4d}.
A careful determination of the critical area still needs to
be performed using the double scaling limit.

It is interesting to look at the fermionic sector of large $N$ 
$4d$ QCD.
Chiral symmetry is broken in the $0$c phase and
the chiral condensate is found to be~\cite{Narayanan:2004cp}
\be
\frac{1}{N}\langle\bar\psi\psi\rangle^{\overline{MS}}(2{\rm GeV}) \approx (174{\rm MeV})^3
\ee
Assuming $N=3$ is large enough, we get
$\langle\bar\psi\psi\rangle^{\overline{MS}}(2{\rm GeV}) 
\approx (251{\rm MeV})^3$ for SU(3).
$m_\pi^2 \propto
m_q$ as expected and~\cite{Narayanan:2005gh}
\be
\frac{f_\pi}{\sqrt{N}}  \approx 71{\rm MeV}.
\ee
This translates to $f_\pi=123$~MeV for SU(3).
This is the first instance we know of where Montecarlo simulations
have indicated large
$1/N$ corrections.
Pseudoscalar masses as well as vector meson masses
were recently computed~\cite{Bali:2007kt}
for $N=2,3,4,6$ and extrapolated
to the large $N$ limit.
It would be interesting to study
current correlators and compute vector meson masses
directly in the large $N$ limit using quenched
momentum techniques.
It would also be interesting to
study the correlations of Dirac eigenvalues as a function of
force-fed momentum and their relation to $f_\pi$
using recent ideas from random matrix theory~\cite{Akemann:2007wf}.

Chiral symmetry is restored in the $1$c phase and
it is a first order 
transition~\cite{Narayanan:2006sd,Narayanan:2006ek}.
The behavior of the Dirac spectrum in the $3$c and
$4$c phases would shed some useful insight into 
dimensionally reduced theories.

\subsection{Twisted Eguchi-Kawai formalism}

Twisted Eguchi-Kawai was originally proposed as a way
to preserve the $U^4(1)$ symmetries on a $1^4$ lattice
all the way to the continuum limit~\cite{GonzalezArroyo:1982hz}. 
The basic idea is to modify the 
gauge action on the $1^4$ lattice
to 
\be
S_{\rm TEK} = -bN\sum_{\mu\ne \nu}^4 {\rm Tr}
\left( 1 - e^{-i\frac{2\pi}{N} n_{\mu\nu}}
U_\mu U_\nu U^\dagger_\mu U^\dagger_\nu\right).
\ee
The twist factors, $n_{\mu\nu}=-n_{\nu\mu}$ are integers.
If one chooses, $n_{\mu\nu}=L;\ \ \mu>\nu$,
and $N=L^2$, then the theory behaves as if it is on a
$L^4$ lattice. This theory has a $Z_N^4$ symmetry
like the usual Eguchi-Kawai model. But this symmetry
is not broken in the strong coupling limit or weak
coupling limit for the above choice of twists. 

There has been a recent revival of the twisted Eguchi-Kawai
formalism of large $N$ $4d$ QCD with the aim of numerically
studying the status of the $Z_N^4$ symmetry as a function
of $b$ from $b=0$ to $b=\infty$.
A surprising new result is
that
twisted Eguchi-Kawai models on $1^4$ lattice
seem to break the $Z^4(N)$ symmetries for large
enough $N$ for a certain range of couplings~\cite{Teper:2006sp}.
One sees a cascade of transitions 
where one goes from $Z_N^{4(r)}\to Z_N^3 \to Z_N^2 \to Z_N^1 
\to Z_N^0 \to Z_M^{4(t)}$. $Z_N^{4(r)}$ and $Z_N^{4(t)}$
corresponds to the fully symmetric phase at $b=0$ and
$b=\infty$ respectively.
The cascade of transitions do not occur for small $N$
($N < 81$) and begin to occur as bifurcations for $N > 81$.
These transitions seem to be strongly first order.
For large enough coupling at a fixed $N$, all the
$Z^4(N)$ are most likely restored. It is hard to see
the restoration numerically starting from the fully
broken phase and this is also the reason why one cannot
confirm if there is 
a reverse cascade that takes it from $Z_N^0 \to Z_N^{4(t)}$.
The above result has been confirmed independently
by Ishikawa~\cite{Ishikawa} and Okawa~\cite{Okawa}.
The dependence on the critical coupling for these
transitions as a function of $L$ has not been studied.
The possibility to define a continuum limit of large $N$
QCD using twisted Eguchi-Kawai model will depend on how
the critical coupling scales with $L$.
Also, other twists, perhaps with prime numbers for $N$
may show a different picture.

\section{Theta parameter in large $N$ $4d$ QCD}
In the instanton dilute gas approximation, the
free energy as a function of the
$\theta$ parameter is given by~\cite{Schafer:1996wv}
\be
f(\theta) = \chi \left(1-\cos\theta\right)
\ee
where $\chi$ is the topological susceptibility which is
expected to go to zero exponentially in $N$.
Large $N$ arguments suggest that $\theta/N$ is the
parameter to keep fixed as $N\to\infty$. 
Since $f(\theta)$ should be a periodic
function in $\theta$, $f(\theta)$ cannot be an
analytical function of $\theta$:
\be
f(\theta) = \frac{\chi}{2} \min_k \left(\theta + 2\pi k\right)^2
\ee

Instanton and large $N$ make qualitatively different predictions
for the moments of the topological charge.
Instanton arguments say that the topological
susceptibility goes down exponentially with $N$.
Large $N$ arguments give a finite topological
susceptibility in the large $N$ limit.
The ratio of the fourth moment to the second moment
would be unity for instantons and
would be zero by large $N$ argument. Large $N$ predicts
$$\lim_{\epsilon\rightarrow\ 0}
<Q>|_{\pi-\epsilon} \ne  
\lim_{\epsilon\rightarrow\ 0}<Q>_{\pi+\epsilon}.$$

Lattice computations~\cite{Lucini:2004yh} 
of the topological susceptibility
for $N=2,3,4,6,8$ show that the large $N$ limit is finite
and is given by $\frac{\chi^{1/4}}{\sqrt{\sigma}}=0.390(14)$.
A high statistics computation of the topological charge
for SU(3)~\cite{Giusti:2007tu} 
and a field theoretical approach~\cite{D'Elia:2003gr}
show that the ratio of the fourth moment to the
second moment is significantly smaller than unity. Both
results
favor the large $N$ argument.
Gauge theories have
also been studied on the lattice by an expansion
around $\theta=0$~\cite{Del Debbio:2006df,Del Debbio:2007kz}.
These results also are in agreement with the large $N$
predications.
A direct measurement of the non-analyticity at $\theta=\pi$
is difficult.

\section{Principal chiral models in $d=2$}

Two dimensional principal chiral models for large $N$ are 
similar to four dimensional large $N$ gauge theories.
The principal chiral model is defined through the 
action~\cite{Campostrini:1994ih}
\be
S=-Nb\sum_{x,\mu}{\rm Tr}\left [ U(x)
\left(U^\dagger(x+\hat\mu)+U^\dagger(x-\hat\mu)\right)\right]
\ee
This model has a global $SU(N)$$\times$$SU(N)$ symmetry
under which
\be
U(x) \rightarrow R^\dagger U(x) L;\ \ \ R,L\in {\rm SU(N)}
\ee
and it undergoes a 
second order phase transition at $b_c=0.3057(3)$.
The theory is in the continuum phase for $b>b_c$ and the
continuum limit is reached by taking
$b\to\infty$.

Define the operator
\be
L(n,m;b) = U(x)U^\dagger(x+n\hat\mu+m\hat\nu);\ \  n\le m .
\ee
It can be used to define the correlation function
\be
G(n,m;b)=\frac{1}{N}\left< Tr L(n,m;b)\right>
\ee
The correlation length is defined as
\be
\xi^2(b) = \frac{1}{4\sin^2\frac{\pi}{L}}
\left[\frac{\tilde G(0,0;b)}{\tilde G(0,1;b)}-1\right]
\ee
where
$\tilde G(p_1,p_2;b)$ is the lattice Fourier transform of
$G(n,m;b)$.
The correlation length diverges at $b=b_c$ and also as
$b\to\infty$.

Like in the case of large $N$ gauge theories, it
is useful to consider the
eigenvalues of $L(n,m;b)$. These eigenvalues are invariant under
the global transformations. Let $\rho(\theta;n,m;b)$ 
define the distribution of eigenvalues.
$L(0,1;b)$  appears in the action and
is analogous to the plaquette operator in large $N$ gauge theories.
$\rho(\theta;0,1;b)$ does not have a gap for $b < b_c$
and has a gap for $b > b_c$~\cite{Campostrini:1994ih}. 
The universal behavior has not yet been analyzed.

\begin{figure}
\vspace{1.1cm}
\begin{center}
\includegraphics[width=0.6\textwidth]{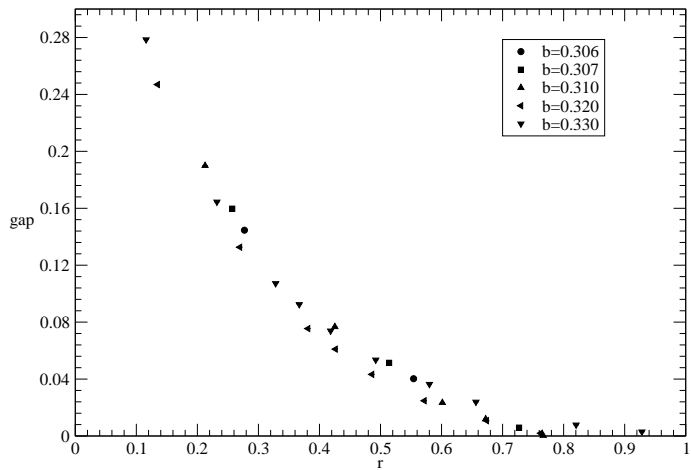}
\caption{Gap as a function of $r$ in the distribution
of $\rho(\theta;r)$\label{gapvsnc}}
\end{center}
\end{figure}

$L(n,m;b)$ is analogous to a Wilson loop operator in
large $n$ gauge theories with $r=\frac{\sqrt{n^2+m^2}}{\xi(b)}$
being the physical length.
We expect
$\rho(\theta;r)$ in the continuum limit
to show critical behavior
such that it has a gap for $r< r_c$ and it does not have
a gap for $r>r_c$. An initial investigation~\cite{vicari} of the gap
as defined in section~\ref{wilcr}
is plotted as a function of $r$ in Fig.~\ref{gapvsnc}
and it suggests the expected picture.
But, a closer look indicates a drift in the critical value
of $r$ as one gets closer to the continuum limit. This
might be an effect of not using smeared $U(x)$. A proper
investigation will have to use smeared $U(x)$ and
the critical size will have be studied as a function
of the smearing factor.

\section{Large $N$ gauge theories with adjoint matter}
Large $N$ gauge theories 
on a $d$ dimensional torus with $p$ adjoint matter fields
can be viewed as a $(d+p)$ dimensional large $N$ gauge
theory in the $p$c phase where the length of the broken
directions are taken to be zero\cite{Aharony:2005bq,Aharony:2005ew}.
Let the masses of the $p$ scalar fields  be the same
and let the
lengths of the periodic directions of the $d$ dimensional
torus be free parameters of the theory.
The Polyakov loops associated with the gauge fields
on the $d$ dimensional torus serve as order parameters.
Such theories in $d=1$ and $d=2$ can be analyzed
pseudo-analytically (with a little bit of numerical help) 
for some region of the parameter space.

The single Polyakov loop in $d=1$ breaks if $p>1$.
Pseudo-analytical analysis shows that there is a line
in the two dimensional coupling constant plane (mass and
length of the circle) that separates the broken phase
from the unbroken phase. 
This is consistent with the
existence of $2$c to $3$c phase transition in $d=3$
and a $3$c to $4$c transition in $d=4$ for 
large $N$ gauge theories on a $d$ dimensional torus.

There are two Polyakov loops and $p\ge 1$.
It has only been possible to perform a large mass (of
the adjoint scalar field) analysis.
The large mass analysis predicts three phases:
(i) Both loops are not broken; (ii) One of the loops
is broken; (iii) Both loops are broken.
This result is again consistent with the
existence of $1{\rm c}\rightarrow 2{\rm c}\rightarrow 3{\rm c}$
cascading phase transition in $d=3$
and a $2{\rm c}\rightarrow 3{\rm c}\rightarrow 4{\rm c}$
cascading transition in $d=4$ for 
large $N$ gauge theories on a $d$ dimensional torus.

\section{Gregory-Laflamme transition}

Consider a $d$ dimensional theory of gravity with no
matter fields and consider the background spacetime to be
${\cal R}^{d-n}\times T^n$. Assume the torus to be
of the same length, $L$, in all directions. 
A $p$-brane solution is a {\sl Schwarzschild black brane}
that is independent of $p$ of the $n$ directions on the torus.
$p=0$ is a black hole.
The entropy of the black brane defines a temperature, $T_H$,
and $t=T_H L$ defines the length of the torus in terms of the
black brane temperature.
There exists a $t=t_{\rm GL(p)}$ such that $p$-brane
decays into a $(p-1)$-brane  as $t$ increases through
$t_{\rm GL(p)}$ and this referred to
as the Gregory-Laflamme transition~\cite{Gregory:1993vy}.
$t_{\rm GL(p)} < t_{\rm GL(p-1)}$ and there
exists a $t_{\rm C(p-1)}$ such that
$t_{\rm GL(p)} < t_{\rm C(p-1)} < t_{\rm GL(p-1)}$.
The free energy for a $(p-1)$ brane is favored
to a $p$ brane as $t$ increases through $t_{\rm C(p-1)}$.
This cascade of transitions is like the
$0{\rm c}\rightarrow 1{\rm c}\rightarrow\cdots\rightarrow d{\rm c}$
cascade observed in large $N$ gauge theories~\cite{Hanada:2007wn}.
There is a relation between these two transitions --
Fermions can be discarded in super Yang-Mills at high
temperatures since they obey anti-periodic boundary conditions
and the theory reduces to Yang-Mills with adjoint scalars.

\section{Other related topics}
There are several other recent developments in the area of
a large $N$ gauge theories that were not presented due
to time constraints.

Several papers considered the case of fermionic matter in two-index
representations,
adding order $N^2$ degrees of freedom and consequently
changing the large $N$ dynamics of the
pure gauge field~\cite{Armoni:2007vb,Kovtun:2007py,Unsal:2007fb}. 
In particular, the extra repulsion between the
eigenvalues of Polyakov loops the matters fields perturbatively generate
can delay or remove the bulk transitions of the pure gauge system. In
one case there is an argument for the absence of all bulk transitions,
indicating volume independence down to
zero size in the continuum~\cite{Kovtun:2007py}.
We do not know of any numerical work testing this prediction.
 
Another topic is the addition of chemical potential for
the quark fields. Here again, fermions will play a
dynamical part and we are not aware of numerical work
pertaining to the large $N$ limit. Some discussion
of the physical implications of the chemical potential
in large $N$ QCD can be found in~\cite{Cohen:2004mw}
and~\cite{McLerran:2007qj}.
\acknowledgments

R. N. acknowledges partial support by the NSF under grant number
PHY-055375. 
H. N. acknowledges partial
support by the DOE under grant number DE-FG02-01ER41165 at Rutgers,
an Alexander von Humboldt award and the hospitality of the Physics
department at Humboldt University, Berlin.

\end{document}